\documentclass[useAMS,usenatbib,onecolumn,usegraphicx]{mn2e}
\usepackage[dvips]{color}
\title[Thermomagnetic instability in hot discs]{Thermomagnetic
instability in hot discs}
\author[Edward Liverts, Michael Mond, and  Vadim Urpin]{Edward Liverts,\thanks{E-mail:
eliverts@bgu.ac.il (EL); mond@bgu.ac.il (MM);  Vadim.Urpin@uv.es (VU)}  Michael Mond, and  Vadim Urpin\\
Department of Mechanical Engineering,  Ben-Gurion
University of the Negev, \\ P.O. Box 653, Beer-Sheva 84105,
Israel\\
A.F. Ioffe Institute of Physics and Technology and Isaac Newton Institute of
Chile, Branch in St.Petersburg, 194021 St. Petersburg, Russia}
\begin{document}

\date{Accepted ---. Received ----; in original form ----}

\pagerange{\pageref{firstpage}--\pageref{lastpage}} \pubyear{}

\maketitle

\label{firstpage}

\begin{abstract}
A linear stability analysis of ionized discs with a temperature gradient and
an external axial magnetic field is presented. It is shown that both
hydromagnetic and thermomagnetic effects can lead to the amplification of
waves and make discs unstable. The conditions under which the instabilities
grow are found and the characteristic growth rate is calculated. The regimes
at which both the thermomagnetic and magnetorotational instabilities
can operate are discussed.
\end{abstract}

\begin{keywords}
accretion, accretion discs, MHD, MRI.
\end{keywords}

\section{Introduction}

The origin of turbulence in astrophysical discs is often attributed to
hydrodynamic and hydromagnetic instabilities that can occur in differentially
rotating stratified gas. The magnetorotational instability (MRI), first investigated
by \cite{vel} and \cite{chandra} and later fully recognized by \cite{bh},
is usually considered as one of the possible candidates to generate such turbulence
and is thought to play an important role in the evolution and dynamics of
astrophysical accretion discs. The growth of the instability
in weakly ionized magnetized discs is of interest for models of star
formation and the subsequent evolution of protostellar discs. Numerical
simulations of the MRI in accretion discs \citep{haw, brand,mats,tork,arlt}
show that turbulence generated can enhance essentially the angular momentum
transport. The MRI has been studied in detail for both stellar and accretion
disc conditions (see, e.g., \cite{fricke,safronov,ach,bh,kai,zhang}). In fact,
the MRI can occur only in a relatively weak magnetic field, but a sufficiently
strong field can suppress the instability completely \citep{bh}. This is
related to the fact that the MRI is basically a long wavelength phenomena
in the sense that it is stabilized for wavelengths shorter than $\lambda_{cr}
\sim 2 \pi c_{A}/ \Omega$ where $c_{A}$ is the Alfven velocity and $\Omega$
is the angular velocity. Defining $\beta = c_s^2/c_{A}^2$ where $c_s$ is the
sound speed, we can estimate that the critical wavelength $\lambda_{cr}$ is
longer than the half-thickness of the disc, $H \sim c_s/ \Omega$ for $\beta >
1$. More rigorous calculations for high and low values of the radial wave
number $k_r$ may be found in \cite{ck}, and \cite{lm}, respectively. In
particular, it has been shown by \cite{lm} that the number of unstable MRI
modes is decreasing with $\beta$. For example, it was found that there exist
only three unstable modes in the disc for $\beta \sim 10^3$. It is the goal
of the present paper to study the effect of the temperature gradient on
stability of magnetized discs in the case of high beta. It will be shown that
the range of unstable wavelengths is significantly widened due to the vertical
temperature gradients and can extend to values shorter than the disc thickness
even in the case $\beta > 1$.

In discs, the centrifugal force almost is balanced by the gravitational
force, while the vertical structure approximately is determined by
hydrostatic equilibrium. The asymptotic analysis [see, e.g.,\cite{regev},
\cite{kk}, \cite{ogilvie}, \cite{sm}] reveals that the vertical temperature
gradient is comparable to that of the density in the case of a small aspect
ratio. The destabilizing effect of a temperature gradient on Alfv\'{e}n waves
has been studies first by \cite{gle,gg} and later on by \cite{dolgurp,urp81}
for plasma in hydrostatic equilibrium. It has been shown that the
thermomagnetic instability (TMI) can arise that transforms a fraction of the
thermal energy into magnetic one. Also, \cite{coppi} has shown that the
vertical temperature gradient in discs combined with the radial rotation shear
give rise to vertically localized ballooning instabilities, the growth rate
of which increases with the temperature gradient. Thus, for a super
adiabatic temperature gradient that can be obtained due to a strong heating
source around the equatorial plane, the growth rate of the ensuing
ballooning modes is found to be comparable to that of the "cylindrical" MRIs.

In this paper, we consider the stability of hot magnetized discs taking into
account the temperature gradient. We will take a look at the MRI and TMI in
various ranges of the wavelength and determine the domains in a parameter
space where each of those instabilities is dominant.

\section{Basic equations}

The dynamics of a magnetized disc is governed by the momentum, continuity, and
induction equations:
\begin{equation}
\rho(\frac{\partial\vec{v}}{\partial t}+\vec{v}\cdot\vec{\nabla} \vec{v})=-\vec{\nabla}p+\vec{J}\times\vec{B}+\rho\vec{G},
\label{mom}
\end{equation}
\begin{equation}
\frac{\partial \rho}{\partial t}+\vec{\nabla}\cdot(\rho\vec{v})=0,
\label{contin}
\end{equation}
\begin{equation}
\frac{\partial\vec{B}}{\partial t}=-\vec{\nabla}\times\vec{E},
\label{ind}
\end{equation}
where $\vec{G}$ is gravity.

The current density that is given by Amp\'{e}re's law:
\begin{equation}
\vec{\nabla}\times\vec{B}=\mu_0\vec{J},
\label{amp}
\end{equation}
where $\mu_0$ is magnetic constant and $\vec{B}$ satisfies a divergence-free condition
\begin{equation}
\vec{\nabla}\cdot\vec{B}=0,
\end{equation}

These set of equations should be complemented by the generalized Ohm's law
that expresses the electric field $\vec{E}$ in terms of the electric
current $\vec{J}$ and gradients of the thermodynamic quantities. It should be
noted that due to quasi-neutrality of plasma the independent thermodynamic
variables are only the temperature and pressure. Then, the Ohm's law reads
(see, e.g., \cite{brag})
\begin{equation}
\vec{E}=
-\vec{v}\times\vec{B}-\frac{1}{qn_e}\vec{\nabla} P_e+
\hat{\eta} \cdot \vec{j}
+\hat{\Lambda}\cdot \vec{\nabla}T,
\label{ohm}
\end{equation}
where the last two terms represent the galvanomagnetic and thermomagnetic
effects, correspondingly. The tensors $\hat{\eta}$ and $\hat{\Lambda}$
depend on the magnetic field $\vec{B}$ and, as a result, the transport
properties of plasma are anisotropic with substantially different properties
along and across the magnetic field if the field is sufficiently strong.
The effect of the magnetic field on the transport properties is characterized
by the magnetization parameter $a_{\rm e} = \omega_{B} \tau$ where $\omega_{B}
= q B /m_{\rm e}$ is the gyrofrequency of the electrons and $\tau$ is their
relaxation time (see, e.g., \cite{spit}). Even poorly conducting protostellar
discs can be strongly magnetized if
the electrons are the main charge carriers \citep{wardle}.

The tensor terms in Eq.(6) can be written as follows
\begin{eqnarray}
\hat{\eta}\vec{J}=\eta\vec{J}+\eta_1(\vec{J}\times\vec{B})+\eta_2\vec{B}(\vec{B}\cdot\vec{J}), \label{galvano}\\
\hat{\Lambda}\vec{\nabla}T=\Lambda\vec{\nabla}T+\Lambda_1(\vec{\nabla}T\times\vec{B})+
\Lambda_2\vec{B}(\vec{B}\cdot\vec{\nabla}T),
\label{thermo}
\end{eqnarray}
where the second terms in both expressions represent the Hall and
the Nernst effects, correspondingly. The coefficients of these expressions
can be obtained from the relations given by \cite{brag,Landau}.
It is convenient to represent the effects of the temperature
gradients by introducing two quantities that have the dimension of velocity:
\begin{eqnarray}
\vec{u}_{1T}&=&\Lambda_1\vec{\nabla}T \\
u_{2T}&=&\Lambda_2(\vec{B}\cdot\vec{\nabla}T)
\label{thermovelocities}
\end{eqnarray}
both of which are of the order of $u_{1T}\sim u_{2T}\sim k_B\tau/
m_e|\vec{\nabla}T|$ for moderate values of the magnetization parameter
$a_{\rm e}$ with $k_B$ being the Boltzmann constant.

The basic state of the considered rotating discs is assumed
to be characterized by an angular velocity that depends on the radial
coordinate alone, $\Omega(r)$, and by hydrostatic equilibrium along the rotation axis $z$;
($r; \varphi; z$) are the cylindrical coordinates. Thus, assuming polytropic equation
of state, asymptotic analysis of the steady-state properties of the discs demonstrates
that the vertical temperature gradients are much larger than the radial ones
\citep{regev,kk,sm}. Consequently, as a model problem, in the current calculations
the steady-state temperature is assumed to vary only vertically.
In addition, the background magnetic field is assumed to be a constant parallel to
rotation axis $z$.

As a first step toward studying the TMI it is instructive to estimate
the relative importance of the thermomagnetic and galvanomagnetic terms contribution
to Ohm's law.
For moderate values of the magnetization parameter and wavelength of perturbations
measured by $c_A/\Omega$ (typical axial wave number for the MRI spectrum), assuming
the disc thickness as a characteristic length scale for the temperature gradient
one finds that the thermomagnetic terms in the Ohm's law and the induction equation are much
bigger than dissipative ones arising due to resistivity terms if
\begin{equation}
\frac{k_B T}{m_e c^2} \; \omega_p^2 \tau^2 \gg \sqrt{\beta}
\label{cond}
\end{equation}
where $\omega_p = \sqrt{n_eq^2 /\varepsilon_0
m_e}$ is the plasma frequency with $n_e$ being the number density of electrons
and $\varepsilon_0$ being the electric constant.
This condition can be satisfied for a very hot plasma with arbitrary values of magnetic
field and small or moderate number density. Thus for example, \cite{Boettcher} have
investigated Galactic black hole candidates. Applying their model to GX-339-4 the relevant data is
that the disc temperature is of the order of $10^6$K with number density of the order of $10^{24}m^{-3}$, and
a corona with temperature of the order of $10^8$K. Such range of parameters, with somewhat higher
temperatures, has also been obtained by \cite{Artemova}. According to (\ref{cond}) such conditions are
indeed favorable for the onset of the TMI. The latter is consequently not
suppressed by resistivity so that the galvanomagnetic terms can be dropped from Ohm's law.

Even though condition (\ref{cond}) may be fulfilled in optically thick as well
as in optically thin environments (as was exemplified in the previous paragraph) we restrict the
current discussion to media such that  radiative heat exchange is efficient in that medium, thus
any thermal transport due to the electron conduction may be neglected and the energy
equation may be dropped out. This indeed simplifies the subsequent calculations. To this end one may
refer to optically thick medium where the radiative transport can be described as diffusive heat
transport with a thermal diffusivity that depends on the temperature and therefore a temperature
gradient can be well exist. It should be emphasized though, that the TMI may also be excited in
optically thin discs, however in such environments, the energy equation should be taken into account.

Consider a linear stability of the discs under axisymmetric short wavelength
perturbations that propagate in $z$-direction. The perturbed electric field
$\vec{E}$, magnetic field $\vec{b}$, current density $\vec{j}$ and
hydrodynamic velocity $\vec{v}$ vary in time and space according to
$\exp(ikz-i\omega t)$ so local approximation is used.
Retaining the thermomagnetic terms in Ohm's law (\ref{ohm}), invoking incompressibility that results
in zero pressure perturbations (the latter also means that the perturbations are restricted to Alfv\'{e}n modes)
and assuming Keplerian angular velocity $\Omega(r)$
one can reduce linearized Eqs.(\ref{mom})-(\ref{ohm}) to the following system of equations:

\begin{equation}
\left [\left (
\begin{array}{cc}
\omega^2+3\Omega^2-\omega^2_A& -2i\Omega\omega\\
2i\Omega\omega& \omega^2-\omega^2_A
\end{array} \right )-
\left (
\begin{array}{cc}
-\omega\omega_{1T}-2i\Omega\omega_{2T}&-\omega\omega_{2T}+2i\Omega\omega_{1T}\\
\omega\omega_{2T}-\frac{1}{2}i\Omega\omega_{1T}& -\omega\omega_{1T}-\frac{1}{2}i
\Omega\omega_{2T}
\end{array} \right )\right ] \vec{b}=0.
\label{ind3}
\end{equation}%
Here the characteristic frequencies are the Alfv\'en $\omega_A=kc_A$ and the
thermomagnetic $\omega_{1,2T}=ku_{1,2T}$ frequencies.

Before turning to the stability analysis it is instructive to
notice that for long wave lengths perturbations
all the matrix elements of second matrix in square brackets
of the last system of equations are small and consequently the
thermomagnetic effects can be neglected. In that case the MRI is
the dominant mode of instability. We will elaborate on that point
in the next section. The main interest of the current work
however, is the case for which the matrix elements of second matrix
in square brackets of the equations (\ref{ind3}) is comparable to the
first one. This happens when: $\omega_{1,2T}\sim \Omega$,
that indeed could occur in hot discs.

\section{Dispersion equation for thermomagnetic waves}

The dispersion relation that is obtained from Eqs.(\ref{ind3}) reads
\begin{equation}
\omega^4+a_3\omega^3+a_2\omega^2+a_1\omega+a_0=0,
\label{disp}
\end{equation}
where
\begin{eqnarray}
a_3&=&2\omega_{1T}\\
a_2&=&-\Omega^2-2\omega_A^2+\omega_{1T}^2+\omega_{2T}^2-i\frac{3}{2}\Omega\omega_{2T}\\
a_1&=&-2\omega_{1T}(\Omega^2+\omega_A^2)\\
a_0&=&-3\Omega^2\omega_A^2+\omega_A^4-\Omega^2(\omega_{1T}^2+\omega_{2T}^2)
+\frac{i}{2}\Omega\omega_{2T}(3\Omega^2-5\omega_A^2).
\end{eqnarray}

If $\Omega=0$, Eq. (\ref{disp}) reduces to the dispersion relation for
Alfv\'{e}n waves modified by the temperature gradients (see \cite{gle,gg,
dolgurp,urp81}). If $\Omega\neq 0$ but $\nabla T=0$, Eq.(\ref{disp})
reduces to the dispersion relation for the MRI obtained by \cite{vel},
and \cite{bh}, which describes the combined effect of inertial and Alfv\'{e}n
waves propagating in rotating discs.

It is instructive to examine a regime when the both MRI and TMI may
operate. This implies the limit of thick disc ($\beta\gg 1$) and $\omega_{1,2T}\leq\Omega$.
For typical wave number of the MRI spectrum ($\Omega/c_A$) such regime reveals itself
if $\zeta\equiv u_{2T}/c_A \leq 1$. Furthermore one should note that in hot discs with
the values of temperatures and number density quoted above the inequality (\ref{cond})
can be hold for such regime and thus magnetic diffusivity may be neglected.
A solution of the dispersion equation for $\zeta\equiv u_{2T}/c_A =0.2$ is depicted
in Fig. 1.
\begin{figure*}
\includegraphics[width=120mm]{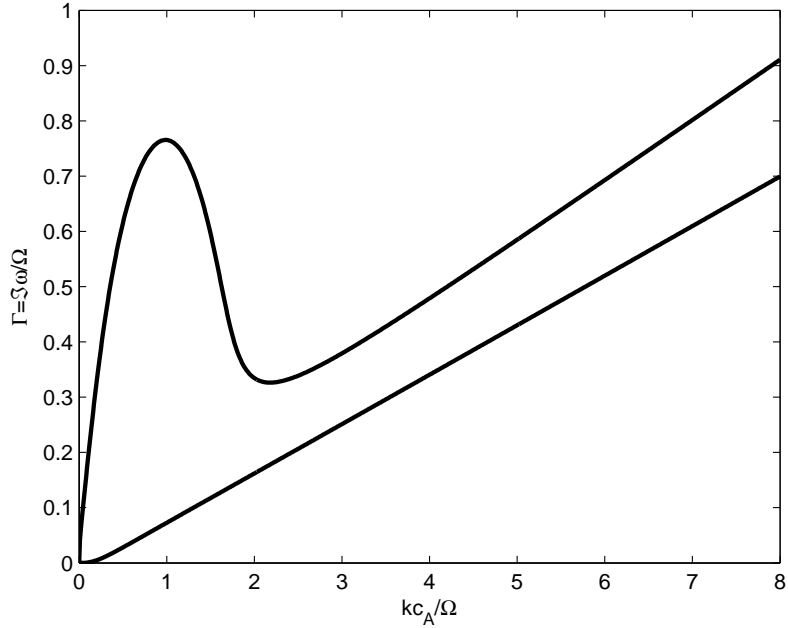}
  \caption{The growth rate, obtained by numerical solutions of eq.(\ref{disp})
for $u_T/c_A=0.2$.}
\label{fig1}
\end{figure*}
It is seen from this figure that there are two modes that become
unstable. The upper curve corresponds to the MRI modified by the temperature
gradient whereas the lower one describes a new instability that occurs due
to the temperature gradient. It can be also obtained from the analytical
consideration
that properties of the instability are qualitatively different in the limiting
cases of $\omega_A \leq \Omega$ (relatively long wavelengths) and $\omega_A
\geq \Omega$ (relatively short wavelengths). In the long wavelength limit
($\omega _A\leq \Omega$), the instability is predominantly the MRI with small
corrections caused by the temperature gradient. In particular, the temperature
gradient is responsible for a finite phase velocity which is vanishing in the
isothermal case. In this limit, the eigenvalue is given by
\begin{equation}
\omega=
\pm (\frac{\Omega}{4}\frac{u_{2T}}{c_A}+i\sqrt{3}\omega_A).
\end{equation}
The corresponding eigenvectors are
\begin{eqnarray}
\vec{b}=\{\frac{2\omega_A}{\sqrt{3}\Omega},\mp 1\}^Tb,\;\;\;
\vec{v}=i\{\mp\frac{2\omega_A}{\Omega},
-\frac{4\omega^2_A}{\sqrt{3}\Omega^2}\}^T\frac{b}{\sqrt{\mu_0\rho}},
\end{eqnarray}
These eigenvectors represent the pair of linearly polarized waves
that propagate in the opposite directions along the $z$-axis. The wave
propagating in the positive direction (this direction
has been assumed conventionally by setting a positive value for $\Omega$)
is exponentially growing but the other one is decaying.

The another branch of modes in the limit of relatively long wavelengths is
characterized by $\omega \sim \Omega$ and small growth rate that is given by
\begin{equation}
\Gamma=Im \omega \approx \pm\frac{5}{4}\omega_{2T}\frac{\omega^2_A}{\Omega^2}
\end{equation}
The eigenvalue and eigenvector for that branch are
\begin{equation}
\omega=\pm(\Omega+i\frac{5}{4}\omega_{2T}\frac{\omega^2_A}{\Omega^2}),\;\;\;
\vec{b}=\{1,\mp 2i\}^Tb,\;\;\;\vec{v}=\{1,
\frac{i}{2}\}^T\frac{b}{\sqrt{\mu_0\rho}}.
\end{equation}
These eigenvectors describe the pair of left elliptically polarized waves that
propagate in the opposite directions. In the absence of temperature gradients
both waves are stable and represent inertial waves of the pure hydrodynamic
nature.

In the limit of relatively short wavelengths (at $\omega \sim \omega_A \geq
\Omega$) and for large temperature gradients such as $u_T/c_A\leq 1$ the
thermomagnetic effects play an important role in the development of
instability. The Alfv\'en waves can be significantly modified by the
temperature gradient and become unstable beyond the stability limit of the
classical MRIs. The growth rate in this regime is given by
\begin{equation}
\Gamma=\Im\omega\approx \frac{\omega_{2T}}{2}(1\pm\frac{2\omega_{1T}+\Omega}{4\omega_A})
\label{gr}
\end{equation}

Finally, in the short wavelength limit and under the condition
$u_T/c_A\leq 1$ the rotation can be neglected and the problem reduces to
that described by \cite{gg}.

In order to gain deeper understanding into the nature of the TMI in the
short wavelength limit, the dispersion equation is rewritten in terms of
$b^{\pm}\equiv b_x\pm i b_y$ in the limit $\Omega \rightarrow 0$. Then,
the dispersion equation reads
\begin{equation}
\left [\omega ^2 +\omega(\omega_{1T}\mp i\omega_{2T} )-\omega^2_A\right
]b^{\pm}=0 \label{circular}
\end{equation}
This equation describes four circularly polarized waves. One pair of waves
propagates along the magnetic field, and its frequency is
\begin{equation}
\omega=\omega_A -\frac{\omega_{1T}}{2}\pm i \frac{\omega_{2T}}{2}.
\label{rcev}
\end{equation}
The corresponding eigenvectors are
\begin{equation}
b^\pm=b,\;\;\;b^\mp=0,\;\;\;\frac{v^\pm}{c_A}=\left [-1-\frac{\omega_{1T}}{2\omega_A}
\pm i\frac{\omega_{2T}}{2\omega_A} \right
 ]\frac{b}{B_0},\;\;\;v^\mp=0.
\label{rcevec}
\end{equation}
The unstable wave (upper sign) is right circularly polarized while the damped wave
(lower sign) is left circularly polarized.

The wave for which the phase shift between perturbations of the magnetic
field and velocity is smaller than $\pi$ ($\pi$ is the appropriate value for
pure Alfv\'{e}n waves) grows exponentially. Another wave for which the phase
shift between the magnetic and velocity perturbations is larger than $\pi$
is damped. The other pair also comprises of two right and left polarized waves that
propagate in the opposite direction to the magnetic field. One of these waves
grows exponentially whereas the other is damped in the same manner as in the
first pair. Summarizing, right circularly polarized waves
propagating in the opposite directions along the magnetic field grow
exponentially with the growth rate
\begin{equation}
\Gamma = \frac{\omega_{2T}}{2}
\end{equation}

The picture of circularly polarized unstable waves allows an additional
interpretation of the instability. To that end it should be noted that
$j^{*\pm}=\mp k b^{*\pm}/\mu_0$ where $j^*$ is the complex conjugate amplitude of
the current density of the circularly polarized waves. The thermomagnetic
contribution to the perturbed electric field is anti Hermitian as it can be
seen from Eq.(\ref{rcevec}). Then, it is easy to show that the average work
over the wave period is given by
\begin{equation}
\langle\vec{E}\cdot\vec{j}\rangle=-\frac{\omega_{2T}}{2\mu_0}|b|^2.
\label{work}
\end{equation}
The time derivative of the average energy $\langle W\rangle$ for the wave
propagating along the external magnetic field is given by
\begin{equation}
\frac{\partial\langle W \rangle}{\partial t}=
\frac{\partial}{\partial t}\int\langle\rho
\frac{v^2}{2}+\frac{b^2}{2\mu_0}\rangle d^3r=
-\int\langle\vec{j}\cdot\vec{E}\rangle d^3r.
\end{equation}
This equation together with Eq. (\ref{work}) clearly shows that the wave
energy grows exponentially with the rate $2\Gamma$.

\section{Summary}

We have considered the thermomagnetic instability in ionized non-isothermal
Keplerian discs that are threaded by an external magnetic field. It has been
shown that, in the short wavelength limit, the temperature gradient allows to
develop the thermomagnetorotational instability for wavelengths satisfying
the condition  $\omega_A \gg \Omega$ for which the MRI does not occurs.
The growth rate of this instability can be comparable to that of the MRI.
Physically, the thermomagnetorotational instability transforms a fraction of
the thermal energy carried out by the heat flux into the magnetic energy of
Alfv\'en waves (modified by the temperature gradient, see \cite{gle,gg,
dolgurp,urp81}). The presence of shear, however, changes substantially the
growth rate of instability. The considered combined magnetorotational and
thermomagnetic instability can play an important role in dynamics of
astrophysical discs.

\end{document}